\documentclass[aps,floats,superscriptaddress,twocolumn,pre]{revtex4-1}
\usepackage{amssymb,amsmath}
\usepackage{amsmath,amssymb}
\usepackage{graphicx}
\usepackage{psfrag,textcomp}
\usepackage{epstopdf}

\def\beq{\begin{equation}}
\def\eeq{\end{equation}}
\def\bea{\begin{eqnarray}}
\def\eea{\end{eqnarray}}
\begin{document}
\title{Thermal fluctuations and stiffening of heterogeneous fluid membranes}
\author{Tirthankar Banerjee}\email{tirthankar.banerjee@saha.ac.in}
\author{Abhik Basu}\email{abhik.basu@saha.ac.in}
\affiliation{Condensed Matter Physics Division, Saha Institute of
Nuclear Physics, Calcutta 700064, India}
\date{\today}

\begin{abstract}
We study the effects of thermal fluctuations on
symmetric tensionless
heterogeneous (two-component) fluid membranes in a simple minimal model.
Close to the critical point $T_c$ of the associated
miscibility phase transition of the composition and for sufficiently strong curvature-composition
interactions, mediated through a composition-dependent bending modulus,
thermal fluctuations lead to enhancement of the effective bending modulus.
Thus, the membrane conformation fluctuations will be {\em
suppressed} near $T_c$, in comparison with a pure fluid membrane,
for which thermal fluctuations are known to reduce the effective bending modulus at all non-zero temperatures.
\end{abstract}

\maketitle

\section{Introduction}\label{intro}
Miscibility phase transitions (MPT) in model heterogeneous membranes are
prominent
examples of phase transitions in two dimensions (2D); see, e.g.,
Ref.~\cite{chamati} for a review on phase transitions with a connection to
biomembranes. Typical experiments on
model lipid bilayers, e.g., artificially prepared lipid bilayers made of lipids
and cholesterol display second order MPTs from
high temperature homogeneous phases to low temperature phase-separated state with coexisting
liquid-ordered ($L_o$) and liquid-disordered
($L_d$) domains \cite{hetero,vet1,sarah-rev} with distinctly different
densities. For instance, a bilayer membrane composed of three components:
dipalmitoylphosphatidylcholine (DPPC),
diphytanoylphosphatidylcholine
(diPhyPC)  and cholesterol clearly display this MPT~\cite{honerkamp}.
  The universal scaling exponents those
characterise MPTs in model lipid bilayers are experimentally found
to be close to those of the two-dimensional (2D) Ising model
\cite{honerkamp,crit}. In general, model lipid bilayers are
symmetric under inversion, due to the identical nature of the two
monolayers, i.e., nothing distinguishes the top from the bottom of
the membranes. While MPTs for the composition field are
well-studied, statistical properties of the membrane conformation
fluctuations at the MPTs have received much less attention
theoretically.

It is well-known that at any finite temperature $T$, a sufficiently large 2D
tensionless
pure or homogeneous (i.e., made of only one lipid) fluid
membrane is {\em always crumpled} due to thermal fluctuations (see
below), i.e., there are no {\em long-ranged orientational
correlations}~\cite{luca,weinbergbook}. As a result, such a membrane
cannot be in a statistically planar configuration beyond a
certain size, determined by $T$ and the microscopic bending
rigidity $\kappa_0$. Physically this is seen as a consequence of the
Mermin-Wagner theorem (MWT)~\cite{mermin}.
At the critical temperature $T_c$ of the
MPT, the presence of the long-ranged composition fluctuations may
introduce long-ranged interactions between different parts of the
membrane. These long-ranged interactions should take the system outside the validity of the MWT. This
leaves the nature of the effective membrane fluctuations open to various
possibilities. Whether or not membrane fluctuations
are affected by the MPTs remains a question of both theoretical and
experimental significance. In this work, we address this issue by introducing a
simple coarse-grained model for a tensionless two-component
heterogeneous membrane, symmetric under inversion, useful as a model for
lipid bilayers with two identical monolayers. We find that at the
physically relevant dimension 2, the behaviour of the membrane at or
near $T_c$ is essentially controlled by the strength of the
interactions between the local curvature and composition
(heterogeneity). If the strength of the interaction coupling between composition
and membrane conformation fluctuations (introduced here through a
composition-dependent bending modulus; see below) exceeds a critical
value that depends on the temperature, the effective large-scale
bending rigidity diverges with the system size $L$. As a result, the
membrane conformation fluctuations are suppressed near $T_c$, or, equivalently,
the membrane appears stiffer. The variance of the associated local normal
fluctuations diverges very slowly with  $L$; on a formal note,
the $L$-dependence is {\em weaker} than the $L$-dependence of the variance of
the orientation fluctuations in 2D classical spin models with continuous
symmetries those display a quasi-long ranged order (QLRO).  On the other hand,
if the interaction
strength is less than the critical value, the effective
scale-dependent bending rigidity vanishes for a sufficiently large
length scale at any $T>0$. Hence the membrane at that scale (or
at larger scales) appears crumpled. In this respect, this behaviour
is qualitatively same as that for a pure fluid membrane. In addition, we
discuss the true area of the membrane and  show that the fractal
dimension $\tilde d$ approaches 2 near $T_c$, if the curvature composition
interaction strength exceeds the critical value, in agreement with the behaviour
of the associated membrane conformation fluctuations. The rest of
the paper is organised as follows: In Sec.~\ref{model}, we construct
our model for a two-component symmetric heterogeneous membrane. In
Sec.~\ref{phase}, we briefly discuss the phase transitions in the
model within a mean-field approach. Then in Sec.~\ref{thermal} we
consider the effects of thermal fluctuations on the membrane
conformations at $T_c$. Next, in Sec.~\ref{genmod}, we discuss a generalised
version of the model above and show that the essential qualitative results do
not change. Finally, in Sec.~\ref{conclu}, we conclude.

\section{The model}\label{model}

For simplicity we ignore the bilayer structure~\cite{lockedmono} and
consider a single membrane consisting of two different components
(lipids) A and B. The local inhomogeneity is appropriately described
by a single composition field $\phi ({\bf x})$, that is defined as
the local difference between the concentrations $n_A$ and $n_B$ of A and
B, respectively: $\phi=n_A-n_B$~\cite{ayton};
$\phi$ is the
order parameter field for the MPT. How composition fluctuations may
affect the membrane conformation fluctuations depends upon the nature of coupling
 between the local composition and curvature. We consider a bilayer with two identical monolayers; for this
 we study a symmetric membrane, invariant under inversion and hence with zero spontaneous curvature. In the simplest description for such a
 membrane, $\phi$ couples with the local mean curvature
 through a composition-dependent bending modulus $\kappa(\phi)$.
  We make a simple choice
  $\kappa(\phi)=\kappa + 2\overline\lambda\phi+2\lambda\phi^2$ giving a
  minimal coupling between the composition and curvature (see, e.g., Ref.~\cite{ayton}
  for a similar model with $\lambda=0$). The sign of $\overline\lambda$ is arbitrary, where as $\lambda>0$ strictly. Therefore, depending
upon the sign of $\overline\lambda$,
  $\kappa (\phi)$ is either higher or lower in A or B rich regions. As a result, domains with smaller
  $\kappa(\phi)$ has a lower free energy cost for supporting a given curvature of either sign. Naturally, regions of
  higher curvature will tend to favour regions of smaller curvature. Given the extensive experimental
evidence in support of second order MPTs in symmetric heterogeneous
membranes, we adopt the standard Ginzburg-Landau free energy
functional for binary
mixtures~\cite{safran,chaikin}, useful near a critical point.
     We study a nearly flat membrane, for which  membrane conformations are conveniently described by a single-valued height field $h(x,y)$ measured from a
perfectly flat base plane in the Monge gauge~\cite{weinbergbook}.
 Obviously, the choice for the base plane is arbitrary. Hence, the membrane
free energy functional density should be invariant under the tilt:
$h\rightarrow h+{\bf a\cdot r}$,
where $\bf a$ is an arbitrary constant three-dimensional (3D) vector and $\bf r$ is a 3D radius vector.

Our model free energy functional $\mathcal F$ for a two-component
tensionless membrane (i.e., zero effective surface tension)
~\cite{surface} to the lowest order in gradients and nonlinearities
is given by
\begin{eqnarray}
 {\mathcal F}&=&\int dS [\frac{1}{2}\kappa(\phi)H^2+\frac{r}{2}
 \phi^2+\frac{1}{2}(\nabla_\alpha\phi)(\nabla^\alpha\phi)
+\frac{u}{4!}\phi^{4}]\nonumber \\&=&\int dS [\frac{1}{2}\kappa
H^{2} +\frac{r}{2}
 \phi^2+\frac{1}{2}(\nabla_\alpha\phi)(\nabla^\alpha\phi)
+\frac{u}{4!}\phi^{4}\nonumber \\ &+&\lambda \phi^{2}
 H^{2} + \overline\lambda\phi H^2],\label{freegen}
\end{eqnarray}
where, $H$ is the mean curvature; $\nabla_\alpha$ is the gradient
operator, $\nabla^\alpha=g^{\alpha\beta}\nabla_\beta$;
$g_{\alpha\beta}$ is the metric on the membrane,
$g_{\alpha\beta}g^{\beta\gamma}=\delta_\alpha^\gamma$~\cite{weinbergbook}.
Further, $\kappa$  is the (bare) bending modulus, $r\sim T- T_c$, $u>0$ is
a coupling
constant, $\lambda,\,\overline\lambda$ are coupling constants which
couple the composition with the mean curvature.  Note that the $\lambda$-term in
(\ref{freegen}), that is quadratic in $\phi$ and $H$, ensures thermodynamic
stability with $\lambda$ chosen to be strictly positive.
With $\phi$ as a number density, taking $\zeta\sim$ microscopic
molecular length ($\sim 10\AA$), $\phi\sim 1/\zeta^2$,
$\lambda\sim K_BT\zeta^4,\,\tilde\lambda\sim K_BT\zeta^2$. Further, surface element
$dS$ is related to the projected surface element $dxdy$ in a flat
reference plane via $dS=dxdy\sqrt g,\,g=det\,g_{\alpha\beta}$.  In
the Monge gauge, the  metric tensor $g_{\alpha\beta}$ is given by
 \[g_{\alpha\beta}=\left(\begin{array}{cc}
 1+(\partial_x h)^2 & \partial_xh\partial_yh\\
 \partial_xh\partial_yh & 1+ (\partial_y h)^2\end{array}\right).\]
In addition, surface element
$dS=dxdy\sqrt{1+({\boldsymbol\nabla}h)^2}$ and mean curvature
 $H= {\boldsymbol\nabla} \left[\frac{- {\boldsymbol\nabla} h}{\sqrt{1+({\boldsymbol\nabla} h)^2}}\right]$ in the Monge gauge. The membrane being
symmetric is invariant under $h \rightarrow -h$. In the Monge gauge,
$dS=
\sqrt{1+({\boldsymbol\nabla} h)^2}dx dy \approx [1+ \frac{1}{2}
({\boldsymbol\nabla} h)^2] dx dy$, assuming small height fluctuations for a nearly flat membrane. Evidently, if the nonlinear forms
of $dS$ and $H$ are included in $\mathcal F$ above, additional
nonlinear terms will be generated. These are {\em geometric
nonlinearities} owing to their origin in the Monge gauge, as opposed
to the thermodynamic nonlinearities in $\mathcal F$.

\section{Phase transition}\label{phase}

It is instructive to begin with a mean field theory (MFT)
description in terms of (assumed constant) order parameter $m=\phi$
and mean curvature $C=-\nabla^2 h$. Neglecting the geometric
nonlinearities and minimising $\mathcal F$ with respect to $m$ and
$C$, we obtain
\begin{eqnarray}
rm+\frac{u}{3!}m^3 + 2\lambda m C^2+\overline\lambda C^2 &=&0,\label{mfm}\\
\kappa C + 2\lambda C m^2+2\overline\lambda mC&=&0 \label{mfc},
\end{eqnarray}
yielding $C=0$ at all $T$ and $rm+\frac{u}{3!}m^3=0$ that describes a
second order transition for the order parameter $m$ with $m=0$ for
$r>0$ ($T>T_c$) and $m^2=-3!r/u$ for $r<0$ ($T < T_c$). Solution $C=0$ is
consistent with the inversion symmetry of the model free energy
(\ref{freegen}), a requirement for any symmetric membrane. Notice that
Eq.~(\ref{mfc}) may be written as
\begin{equation}
\kappa_m C=0,
\end{equation}
where $\kappa_m=\kappa+2\overline\lambda m+2\lambda m^2$ is the effective bending
modulus in MFT ( we ignore the possibility of $\kappa_m=0$ for a
nearly
flat membrane). Clearly, with $\overline\lambda>0$, $k_m$ is reduced in
domains with $m<0$, but enhanced in domains with
$m>0$; $\lambda>0$ contributes positively to $\kappa_m$ for both signs of $m$.
This is consistent with the interpretation of the
$\overline\lambda$- and $\lambda$-terms in $\mathcal F$ as discussed above.

\section{Thermal fluctuations and bending modulus near $T_c$} \label{thermal}

We begin with the partition function given by
\begin{equation}
{\mathcal Z}=\int Dh D\phi exp[-\beta {\mathcal F}],
\end{equation}
$\beta=1/k_BT$. Taking Boltzmann constant $k_B=1$, $T/\kappa$ is a
dimensionless number in the problem. We wish to explore the
possible nearly flat configuration of the membrane, such that the
available thermal energy $\sim k_BT \ll $ bending energy, the scale
of the latter being set by $\kappa$. Hence, we take
$T/\kappa<<1$~\cite{chaikin,luca}. Below we construct a
perturbation theory with $T/\kappa$ as the (small) expansion
parameter; we calculate fluctuation corrections to $\kappa$  to
first order in $T/\kappa$. It is convenient to truncate the free
energy functional (\ref{freegen}) up to $O(T/\kappa)$. We find
 to the lowest order in $T/\kappa$, the free energy $\mathcal F$ is given by
\begin{eqnarray}
{\mathcal F}&=&\int d^2x[\frac{\kappa}{2}(\nabla^2 h)^2 -
\frac{\kappa}{4}(\nabla^2 h)^2 ({\boldsymbol\nabla} h)^2 \nonumber
\\&-&\kappa \nabla^2 h\nabla_\alpha h\nabla_\beta h
\nabla_\alpha\nabla_\beta h + \frac{r}{2}\phi^2
+\frac{1}{2}g^{\alpha\beta}(\nabla_\alpha\phi)(\nabla_\beta\phi)\nonumber
\\&+&\frac{1}{4}({\boldsymbol\nabla}\phi)^2 ({\boldsymbol\nabla}h)^2+
\lambda\phi^2 (\nabla^2 h)^2+  \overline\lambda \phi
(\nabla^2 h)^2 ],\label{freegen1}
\end{eqnarray}
where the nonlinear terms are kept
up to $O(T/\kappa)$~\cite{tkappa}. Notice that (\ref{freegen}) and hence (\ref{freegen1}) is invariant
under inversion of the membrane, i.e., under $h\rightarrow -h$, as it should be for a symmetric membrane.
However, the $\overline\lambda$-term in (\ref{freegen}) or (\ref{freegen1}), being linear in $\phi$, breaks the Ising symmetry. For the special case of
$\overline\lambda=0$, the Ising symmetry of $\phi$ is restored.

Consider a pure (homogeneous) fluid membrane, $\phi=0$. It is evident from (\ref{freegen1}) with $\phi=0$ that if the geometric
nonlinearities are included, then $\kappa$ receives a negative fluctuation
correction proportional to $\langle ({\boldsymbol
\nabla}h)^2\rangle$~\cite{luca}. In a renormalisation group (RG) language
this implies that the effective scale-dependent bending modulus $\kappa(l)$ has no
fixed point
and flows to a negative value for a large length scale at any finite temperature
$T>0$.
This result is interpreted as the impossibility of finding a (statistically) flat
membrane beyond a
finite scale determined by $T$ and the bare or small-scale bending modulus $\kappa_0$ at dimension $d\leq 2$~\cite{luca}. On the
other hand, for a heterogeneous membrane $\phi\neq 0$, and the $\lambda$-term positively
contributes to the scale-dependent $\kappa(l)$ by an amount proportional to
$\langle\phi^2\rangle$.
This contribution formally diverges at $T_c$ and hence can potentially lead to
stiffening of membranes. Thus, for an
inhomogeneous membrane, the competition between the geometric nonlinearity and the
nonlinearity of the curvature-composition
interaction, mediated by a composition-dependent
$\kappa(\phi)$ should determine the membrane fluctuations in the
thermodynamic limit.

We employ the perturbative Wilson RG
procedure~\cite{chaikin,rg-book} to the lowest order in $O(T/\kappa)$ to
circumvent the difficulty in perturbative
expansions near $T_c$ due to the large fluctuations.
 In addition, we
obtain  our results within a harmonic approximation for $\phi$, i.e., we set
$u=0$ for simplicity. We eliminate fields
$h({\bf q})$ and $\phi ({\bf q})$ with wavevector
$\Lambda/b<q<\Lambda,\,b>1,\,\Lambda$ an upper cut off for wavevector, by
integrating over them perturbatively up to the
one-loop order and then rescaling wavectors according to ${\bf q}'=b{\bf
q}$ (or rescale real space coordinate $\bf x$ according to ${\bf
x}'={\bf x}/b$)
 Since we are interested to find renormalisation of
$\kappa$ due to the geometric nonlinearities and the
curvature-heterogeneity couplings, it is convenient to  let $h ({\bf
x})$ rescale as $h({\bf x'})=h({\bf x})/b$ with ${\bf x'}={\bf
x}/b$. Further,  rescale $\phi$ by $\phi ({\bf
x^\prime})=b^{-\epsilon/2}\phi({\bf x})$, $\epsilon=2-d$. Under this rescaling
$\kappa'=b^{-\epsilon}\kappa,\,\lambda'=b^0\lambda,
\,\overline\lambda'=b^{-\epsilon/2}\overline\lambda$.

\begin{figure}[htb]
\includegraphics[width=5cm,height=6cm]{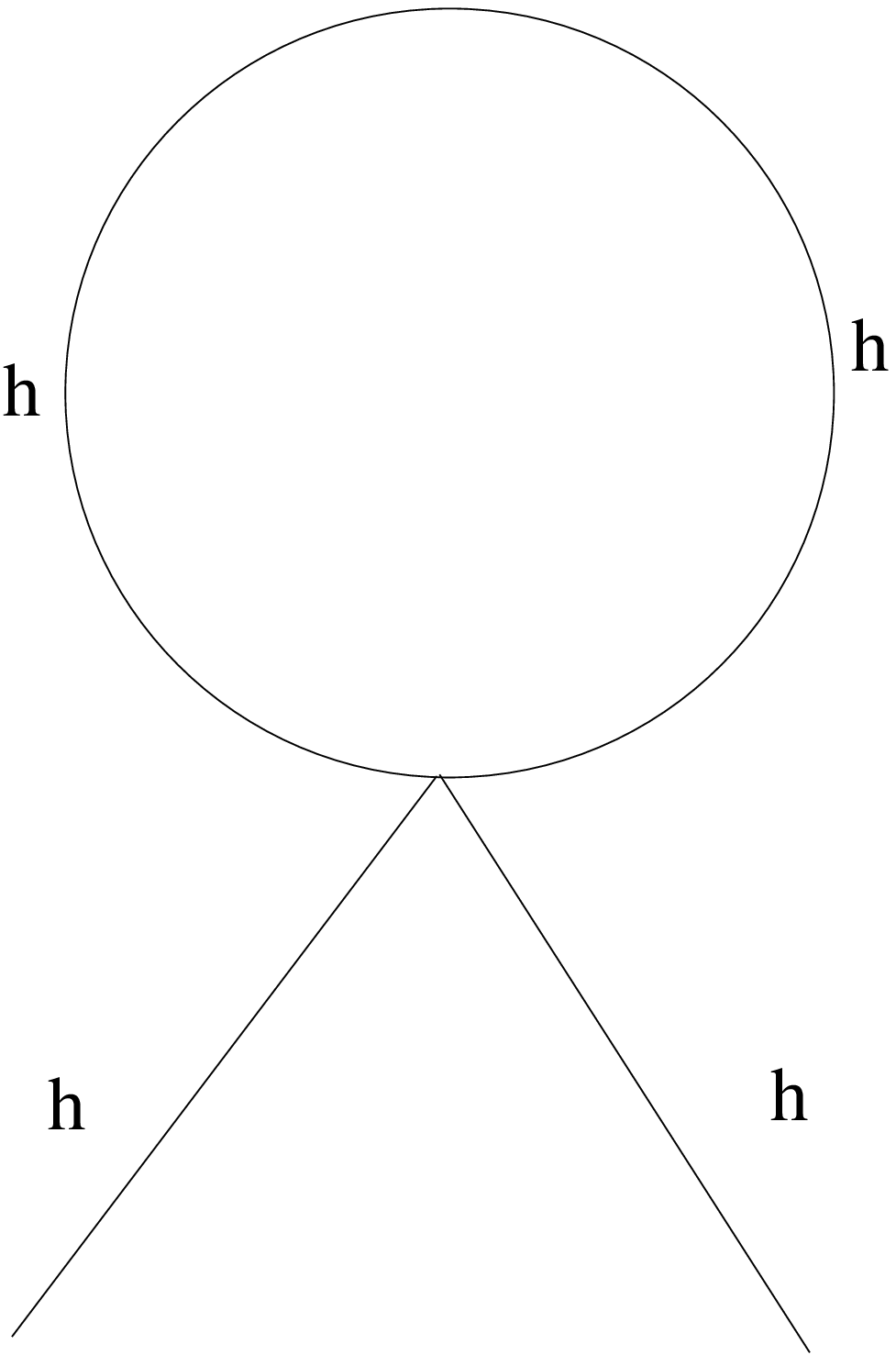}
\caption{One-loop Feynman diagram contributing to the fluctuation correction
to $\kappa$. This originates from the geometric nonlinearity; see
Eq.(\ref{gradh}).}\label{fd1}
\end{figure}

\begin{figure}[htb]
\includegraphics[width=5cm,height=6cm]{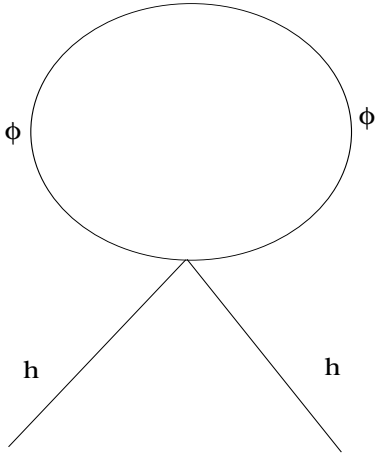}
\caption{One-loop Feynman diagram contributing to the fluctuation correction
to $\kappa$. This is due to the composition fluctuations and exists only for
inhomogeneous membranes; see Eq.(\ref{phi}).}\label{fd2}
\end{figure}

One-loop corrections to $\kappa$ may be straight forwardly extracted
from (\ref{freegen1}) by contracting fields in the nonlinear terms leaving two
external $\nabla^2h$ legs~\cite{renlam}. We follow the procedure as outlined above and find perturbatively
\begin{eqnarray}
 b^\epsilon\kappa^\prime -\kappa &=& -\kappa \frac{3}{2} \langle({\boldsymbol\nabla} h)^2\rangle_\Lambda + 2\lambda \langle\phi^2\rangle_\Lambda\nonumber \\ &-&\overline\lambda^2
 \int_{\Lambda'}^\Lambda\frac{d^2q}{(2\pi)^2}\langle |\phi({\bf q})|^2q^4 |h({\bf q})|^2\rangle,\label{kdiscrete}
\end{eqnarray}
where $\Lambda/\Lambda'=b$ and
\begin{eqnarray}\label{gradh}
 \langle({\boldsymbol\nabla} h)^2\rangle_\Lambda &=&
 \int_{\Lambda/b}^{\Lambda}\frac{d^2q}{(2\pi)^2} \frac{T q^{2}}{\kappa q^4} = T
\frac{ \ln b}{2 \pi \kappa},
\end{eqnarray}
\begin{eqnarray}\label{phi}
 \langle\phi^2\rangle_\Lambda &=& \int_{\Lambda/b}^{\Lambda}
\frac{d^2q}{(2\pi)^2} \frac{T}{q^2}=
  \frac{T\ln b}{2\pi}.
\end{eqnarray}
The two relevant one-loop Feynman diagrams corresponding to the corrections
(\ref{gradh}) and (\ref{phi}) are shown in Fig.~\ref{fd1} and Fig.~\ref{fd2}, respectively.

By using Gaussian decomposition and $\langle |h({\bf q})|^2\rangle = T/\kappa
q^4$,
we note from Eq.~(\ref{kdiscrete}) that the contribution to $\kappa^\prime$
from the $\overline\lambda$-term in (\ref{freegen}) or (\ref{freegen1}) is linear in
$T/\kappa$ where as the remaining contributions are $O(T/\kappa)^0$. Thus,
we neglect this $O(T/\kappa)$ contribution from the $\overline\lambda$-nonlinear term below.

 Now, let $b=e^{dl}\simeq 1+dl$. Then,
 \begin{equation}
  \frac{d\kappa}{dl}=\kappa[-\epsilon - \frac{3T_c}{4\pi\kappa} +\frac{2\lambda T_c}{2\pi\kappa}].\label{kcont}
 \end{equation}
 Notice that in the flow Eq.~(\ref{kcont}), we have set $T=T_c$, since we are at the critical temperature of the MPT.
 At the RG fixed point (FP), $d\kappa/dl=0$ which yields
 \begin{equation}
 -\epsilon - \frac{3T_c}{4\pi} + \frac{2\lambda_c T_c}{2\pi}=0,\label{fp}
 \end{equation}
 which defines a critical $\lambda_c$:
 \begin{equation}
 \lambda_c=\frac{\pi\epsilon}{T_c}+\frac{3}{4},\label{lambdac}
 \end{equation}
  for this unstable RG FP. Thus for $\lambda<\lambda_c$,
 \begin{equation}
 \frac{d\kappa}{dl}<0
 \end{equation}
 implying that under successive applications of the RG procedure, scale-dependent $\kappa(l)$ reduces linearly with
 $l$, eventually becoming zero at a particular scale which depends upon $\kappa$
and $\lambda$~\cite{phiren}; see also
  Ref.~\cite{luca}. Thus
even for a heterogeneous fluid membrane, for a weak
  coupling $\lambda < \lambda_c$, the membrane still
crumples at any $T$ for a sufficiently large size. The discrete recursion
relation for $\kappa$ at 2D is given as
 \begin{equation}
 \kappa=\kappa_0 - \frac{3T_c}{4\pi}\ln b + \frac{\lambda T_c}{\pi}\ln b,
 \end{equation}
 Noting that $\ln b=\ln (\Lambda/q)=-\ln (qa_0)$, where $a_0\sim \Lambda^{-1}$ is a microscopic length, we obtain a $q$-dependent $\kappa(q)$
 \begin{equation}
 \kappa(q)=\kappa_0+\frac{3T_c}{4\pi}\ln (q a_0) -\frac{\lambda T_c}{\pi}\ln (q a_0).\label{kappaq}
 \end{equation}
  Similar to Ref.~\cite{luca}, Eq.~(\ref{kappaq}) allows us to define a de
Gennes-Taupin persistence length $\xi\sim 1/q$, such that $\kappa(\xi)=0$. This yields
  \begin{equation}
  \xi=a_0\exp\left(\frac{4\pi \kappa_0}{T_c(3-4\lambda)}\right).
  \end{equation}
  Clearly, as $\lambda\rightarrow \lambda_{c_-}=3/4$ at 2D, the membrane crumples less and less, and $\xi\rightarrow \infty$. 

 In contrast, for $\lambda>\lambda_c$,
 \begin{equation}
 \frac{d\kappa}{dl}>0
 \end{equation}
 generically and hence, $\kappa$ grows under successive applications of the RG transformations. Therefore, the membrane at larger scale
 should appear {\em stiffer} than it is at smaller scales. Clearly, as dimensionality rises, i.e., as $\epsilon$ reduces and becomes negative through zero,
 $\lambda_c$ reduces. Formally, thus, at higher dimensions, weaker
curvature-heterogeneity interactions through $\kappa(\phi)$ are enough
for $\kappa(l)$ to grow under RG transformations. To know whether or not a
heterogeneous fluid membrane with $\lambda >\lambda_c$ crumples in the
thermodynamic
 limit, we need to find $\kappa (q)$ in the limit $q\rightarrow 0$ with $\lambda>\lambda_c$. Define $\Delta\lambda = \lambda-\lambda_c >0$.
 Then,
 \begin{equation}
\kappa(q)= -\frac{\Delta\lambda T_c}{\pi} \ln (qa_0) + \kappa_0,
\end{equation}
which diverges in the thermodynamic limit $q\rightarrow 0$.

Statistical flatness of a membrane may be ascertained
by the variance of the fluctuation of the local normal ${\bf n}=(-{\boldsymbol
\nabla} h,1)$.
We define fluctuation in the normal,
$\delta {\bf n}= -{\boldsymbol \nabla} h$ as the deviation from the normal for a
perfectly flat surface.
Now calculate $\Delta_0=\langle (\delta {\bf n})^2\rangle$. If $\Delta_0$
at a given temperature for a membrane (pure or inhomogeneous) is independent
of the system size $L$, $\Delta_0$ remains finite even in the thermodynamic
limit $L\rightarrow\infty$. This implies that the membrane at that
temperature possesses a long-ranged orientational order in the thermodynamic
limit, or, the membrane is statistically flat~\cite{chaikin,weinbergbook}.
Else the membrane is crumpled. We find at 2D
\begin{equation}
\langle (\delta {\bf n})^2\rangle=\int \frac{d^2 q}{(2\pi)^2}\langle |\delta {\bf n}_{\bf q}|^2\rangle\sim \int_L\frac{{d^2q}}{(2\pi)^2}\frac{T_c}{\kappa(q) q^2},
\end{equation}
where, a subscript $L$ refers to a lower momentum cutoff $\sim 1/L$.
With $\kappa(q)\sim -\frac{T_c}{\pi}\Delta\lambda\ln (qa_0)$, clearly, within the validity of our calculations ($T_c/\kappa \ll 1$)
\begin{equation}
\Delta_0\sim -\frac{1}{\Delta\lambda}\int\frac{dq}{q\ln (qa_0)}\sim - \frac{1}{\Delta\lambda}\ln\ln (\frac{a_0}{L}) \label{weakL}
\end{equation}
in the critical region ($T=T_c$), regardless of the specific value
of $T_c$, so long as $\lambda>\lambda_c$.{ Thus, $\Delta_0$ still diverges for
$L\rightarrow \infty$, albeit very slowly and is finite for any finite $L$. In fact, the system size $L$ required to have a given $\Delta_0$ is given by $L/a_0\sim \exp (\exp[\Delta_0\Delta\lambda])$. Thus, $L$ rises very rapidly with $\Delta\lambda$}. Compare this with a pure
fluid membrane, for which $\kappa (q)$ vanishes for a low enough
$q$. Thus for a pure fluid membrane $\Delta_0$ should diverge even for a
finite system size ($\sim\xi$).
Notice that since $\kappa(q)$ here remains finite (in fact growing)
even down to $q\rightarrow 0$, $\xi$ is formally infinitely large. As a result, for a finite membrane of any size, $\Delta_0\neq 0$ and hence, a non-zero orientational correlation should be present.
The very weak divergence of $\Delta_0$ with $L$
 means its experimental detection should be rather difficult;  crumpling may be
observed only in a very large membrane. Notice that the $\ln\ln
(L/a_0)$-dependence on $L$ in Eq.~(\ref{weakL}) is
even {\em weaker} than the well-known $\ln L$-dependence of the variance of
orientation fluctuations in 2D classical spin systems having continuous
symmetries which display QLRO~\cite{chaikin}.

\begin{figure}[htb]
\includegraphics[width=8cm,height=7cm]{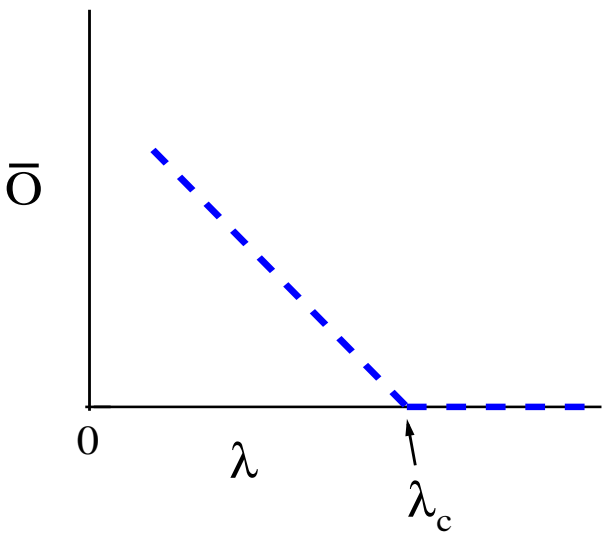}
\caption{(Colour online) Schematic variation of order parameter $\overline O$
as a
function of $\lambda$; $\overline O\propto (\lambda-\lambda_c)$ for
$\lambda\leq \lambda_c$, $\overline O=0$ for $\lambda\lambda_c$ as shown by
the blue broken line; see text.}\label{opm}
\end{figure}

\begin{figure}[htb]
\includegraphics[width=8cm,height=7cm]{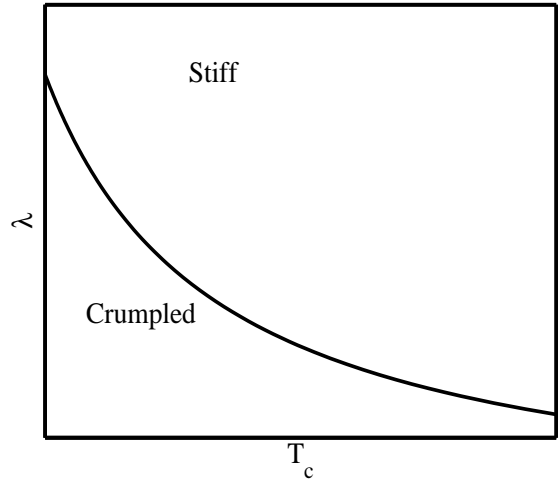}
\caption{Schematic phase diagram in the $\lambda-T_c$ plane  for
$\epsilon>0$. Black continuous line represents
Eq.~(\ref{lambdac}).}\label{phasediag}
\end{figure}

Noting that (\ref{lambdac}) yields an unstable FP for
$d\kappa/dl=0$, our results above may be interpreted as a phase
transition in the membrane between a crumpled phase (finite $\xi$,
$\Delta\lambda <0$) and a stiff phase (diverging $\xi$,
$\Delta\lambda >0$) for a given $T=T_c$ with $\lambda$ appearing as
the control parameter. We take $\overline O = [\ln
(\xi/a_0)]^{-1}=-T_c\Delta\lambda/(\pi\kappa_0)$, with $\lambda<\lambda_c$ and $\overline O=0$ with $\lambda\geq \lambda_c$ as the order
parameter; thus $\overline O$ in the crumpled phase  rises smoothly from zero,
as $\lambda$ is reduced from $\lambda_c$. Thus, with
$\lambda$ as the control parameter, the ``order parameter exponent''
is 1. In the stiff phase, $\overline O$ is naturally zero. We plot $\overline O$ versus $\lambda$ in Fig.~\ref{opm}.
 Figure~\ref{phasediag} shows
a schematic phase diagram in the $\lambda-T_c$ plane.

\begin{figure}[htb]
\includegraphics[width=6cm,height=5cm]{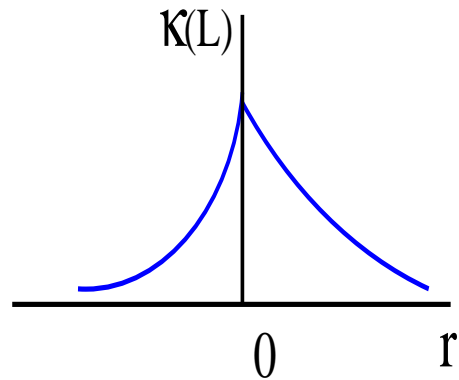}
\caption{(Colour online) Schematic plot of renormalised $\kappa$ for a system
of size $L$ as a function of $r=T-T_c$; rise in $\kappa$ as
$T\rightarrow T_c$ due to a nonzero $\lambda$ is shown.}\label{kappa}
\end{figure}

 Notice that the membrane fluctuations are unaffected by the $\overline\lambda$-term
 to the leading order in $T/\kappa$. These suggest that in the long wavelength limit,
 the Ising symmetry is restored. So far we have discussed the membrane fluctuations
 at the MPT, i.e., at $T=T_c$. Away from $T_c$, contribution of the $\lambda$-term to renormalised $\kappa$ is
 small; renormalised $\kappa$ is dominated by the contributions from
 the geometric nonlinearities. Therefore, at $T\neq T_c$, the large scale properties of a heterogeneous fluid membrane
 described by (\ref{freegen}) is identical to a pure fluid membrane. However,
as one approaches $T_c$ from either side, the contribution from the
$\lambda$-term becomes significant. For a membrane of finite linear size $L$, $\langle
({\boldsymbol\nabla}h)^2\rangle$ scales as $\ln L$, where as $\langle \phi^2\rangle
\sim \ln \frac{L}{L\sqrt r+1}$ at $T>T_c$, i.e., $r>0$. Clearly, the latter
contribution rises as $r\rightarrow 0$, and hence, renormalised $\kappa$, that
now depends on $L$ rises in magnitude. Therefore, any measurements of
$\kappa$ as a function of $T$ should detect this rise as $T_c$ is approached.
Similar behaviour for $\kappa(L)$ follows for $T<T_c$ as well. A schematic plot
of $\kappa(L)$ versus $r$ is shown in Fig.~\ref{kappa}. The general nature of
this plot remains unchanged regardless of whether $\lambda>\lambda_c$ or not.

We now calculate the effective membrane area $S=\langle
\sqrt{1+({\boldsymbol \nabla}h)^2}\rangle S_0$ and compare it with the base area $S_0$
(the projected area on the Monge gauge reference plane) by defining the fractal
dimension $\tilde d$ of the membrane via the relation $S\propto S_0^{\tilde
d/2}$. Hence, for $\tilde d=2$, $S\propto S_0$ corresponding to a membrane
that is on average flat, while for $\tilde d>2$, $S$ rises faster than $S_0$ and
hence a more corrugated membrane at larger scales is obtained. We find at $T_c$, to the lowest order in
$h$-fluctuations~\cite{luca}
\begin{eqnarray}
 \frac{\tilde d}{2}&=&\frac{d\ln S}{d\ln S_0}\simeq
1+\frac{1}{2}\frac{d}{d\ln S_0}\langle ({\boldsymbol
\nabla}h)^2\rangle\nonumber \\
&=&1+\frac{1}{8\pi}\frac{T_c}{\kappa}.
\end{eqnarray}
Thus, if $\lambda<\lambda_c$, renormalised $\kappa$ gets smaller at larger
scales, and hence $\tilde d>2$, implying a highly corrugated membrane at large
scales. In contrast, for $\lambda\geq \lambda_c$, $\tilde d\rightarrow 2$, and
hence a nearly flat membrane is obtained. These conclusions are consistent with
the behaviour of $\langle (\delta {\bf n})^2\rangle$ for $\lambda <\lambda_c$
and $\lambda>\lambda_c$, respectively. We close this Sec. with a technical
comment. The phase transition elucidated above takes place at a
fixed temperature $T_c$, with $\lambda$ appearing as the control parameter.
Thus, experimentally,
various realisations of a heterogeneous membrane described by (1), will exhibit
different characteristic
membrane fluctuations at $T_c$, depending upon whether $\lambda>\lambda_c$ or
not. This is different from
the usual crumpling transition of membranes~\cite{weinbergbook}, where the
membrane goes from the high temperature crumpled phase
to the low temperature flat phase through a critical temperature of the
crumpling transition; the temperature
is the control parameter of the usual crumpling transition, unlike in the
present case.

\section{Generalised Model}\label{genmod}

Our results above depend crucially on the positivity of the $\lambda$-term  in
(\ref{freegen}).
We now generalise (\ref{freegen}) by considering an arbitrary form of
interaction $\tilde f$ between the local curvature and inhomogeneity, maintaining only the
 tilt invariance and invariance
under inversion of $h$. We briefly discuss its effects on the membrane fluctuations near $T_c$. Expanding $\tilde f$ in powers of $\phi$ and
$(\nabla^2 h)^2$, we write the generalised free energy functional ${\mathcal
F}_g$ as
\begin{eqnarray}
&&{\mathcal F}_g
 =\int dS \{\frac{\kappa}{2} (\nabla^2 h)^2 + \frac{r}{2}\phi^2+\frac{1}{2}
g^{\alpha\beta}(\nabla_\alpha \phi) (\nabla_\beta \phi)\nonumber \\&&+
\frac{u}{4!}\phi^4 + \sum_{m,n}A_{m,n}\phi^m (\nabla^2 h)^{2n}\}.
\label{genf}
 \end{eqnarray}
Thus, comparing (\ref{genf}) with (\ref{freegen}) above we find
\begin{equation}
A_{1,1}=\overline\lambda,\,A_{2,1}=\lambda.
\end{equation}
All the other $A$'s are zero in (\ref{freegen}). It is clear that contributions of all terms in $\tilde f$ with $n>1$ to renormalised $\kappa$ are higher order
in $T/\kappa$. Thus, in a perturbation theory first order in $T/\kappa$, all
such terms may be ignored here. The remaining terms in $\tilde f$ are all
quadratic in $\nabla^2 h$:
\begin{equation}
\tilde f=\sum_m A_{m,1}\phi^m (\nabla^2 h)^2.\label{gmod}
\end{equation}
For reasons of thermodynamic stability, the series on the right side
of (\ref{gmod}) must be truncated at an even $m=m_{max}$, with
$A_{m_{max},1}>0$. Notice that under spatial rescaling ${\bf
x'}={\bf x}/b$, $A_{2m}'=b^{(m-1)(d-2)}$, implying that for $m>1$,
all of $A_{2m}$ decreases under the successive application of
rescaling. At $d<2$, following the calculation outlined above it is
straightforward to obtain the flow equation for $\kappa$ for the
generalised model. We obtain at $d=2-\epsilon$ to the leading order
in $T/\kappa$
\begin{equation}
\frac{d\kappa}{dl}=-\epsilon-\frac{3T_c}{4\pi} + 2A_{2,1}\langle
\phi^{2}\rangle, \label{kappgen}
\end{equation}
up to the one-loop order. Equation (\ref{kappgen}) is identical in form to
its counter part (\ref{kcont}) above.
  However, unlike the previous case, $A_2$ (in fact all of $A_{2m}$
with $m<m_{max}$) has one-loop fluctuation corrections at
$O(T/\kappa)^0$. For $d\leq 2$ and at $T=T_c$, fluctuations corrections to $A_{2m}$
diverge with $L$ and will dominate over the corresponding bare
parameters. Since $A_{2m}$ with $m=m_{max}$ is necessarily positive,
it is expected that all the fluctuation-corrected, i.e.,
renormalised or effective $A_{2m}$ should be positive. Then
Eq.~(\ref{kappgen}) with $A_{2,1}$ being interpreted as the
renormalised quantity (hence positive) yields similar behaviour for
$\kappa(q)$ from (\ref{freegen}). In the generalised model $A_{2m}$
with $m=m_{max}$ appears as the tuning parameter~\cite{1storder}.


\section{Summary and outlook}\label{conclu}

We have thus proposed a simple coarse-grained model for symmetric
heterogeneous tensionless membranes in terms of the local mean
curvature and a
suitably defined composition field $\phi$ in the Landau-Ginzburg approach  to study
 the nature of membrane conformation
fluctuations  near the critical point of the MPT. Our model should be useful for studying
symmetric bilayers made of identical monolayers with strong inter-monolayer interactions. Within the range of validity
of a
systematic perturbative expansion in $T/\kappa \ll 1$ (with
$T=T_c$), we show that the lowest order nonlinear term that obeys
the Ising symmetry for $\phi$ leads to an enhancement of
renormalised or effective $\kappa$, provided its strength $\lambda$
exceeds a critical value. In that case renormalised $\kappa$
diverges in the thermodynamic limit, albeit very slowly. Nonetheless
$\Delta_0$, the variance of the local normal fluctuations, grows with
system size $L$, although very weakly, $\Delta_0\sim -\ln\ln (a_0/L)$.
This weaker than $\ln L$-dependence, as typically found for the variance of the
orientation fluctuations in classical 2D spin models with XY-like symmetry with
QLRO. This aspect makes it a theoretically intriguing result.
Thus, the membrane should  appear crumpled in the formal
thermodynamic limit, which in practice may be observed at a very large scale due to the very weak $L$-dependence of $\Delta_0$; experimental studies on finite size membranes may not be able
to observe this. We have argued that our results may be interpreted
as a phase transition between a crumpled phase  (finite $\xi$) and a stiff
phase (diverging $\xi$). With $\lambda$ as a control
parameter for a given $T_c$, the transition is second order
characterised by an order parameter $\overline O=[\ln(\xi/a_0)]^{-1} $. A generalised
model by us predicts qualitatively similar results.  With $T_c\sim 300$K and bare
$\kappa\sim 10^{-12}$ erg~\cite{kappavalue} for a lipid bilayer,
$T_c/\kappa\sim 0.05$ is small. Thus, our results should hold for
MPT in a typical model heterogeneous bilayer. By tuning $\lambda$,
experiments (e.g., flicker experiments) on MPTs in a two-component symmetric model lipid bilayer
should reveal suppression of membrane conformation fluctuations (or,
membrane height fluctuations) for $\lambda
>\lambda_c (T_c)$, as $T_c$ is approached. Hence, our results offer an
experimental route
 to investigate the curvature-composition
interactions
in an inhomogeneous membrane.
Our results depend sensitively on $\lambda$.
Since $\lambda\sim\zeta^4k_BT$, performing experiments on model
heterogeneous membranes with different sizes of the constituent
lipid molecules (and hence varying $\zeta$) should be a promising
route to test our results experimentally.
Our results highlight the significant differences in the  fluctuations of
asymmetric~\cite{asym-mem} and symmetric inhomogeneous membranes near $T_c$.

We close our discussions with some technical comments. While
contributions to renormalised $\kappa$ from the
$\overline\lambda$-term is neglected for small $O(T/\kappa)$, for
$T_c/\kappa$ not very small, these may be retained. Notice that this
contribution reduces $\kappa$, see also Ref.~\cite{ayton}. We have
treated $\phi$ only up to the quadratic order and have set $u=0$.
Effects of a non-zero $u$ may be accounted for by considering $\phi$
as the {\em renormalised composition field} with a renormalised
critical temperature and a non-zero anomalous dimension. To the
lowest order in $u$, the only effect would be to shift $T_c$ and our
basic results should stay unchanged. The universal properties of
$\phi$-fluctuations are described by the Ising universality class by
the $\phi^4$ term. It is well-known that~\cite{chamati} a small
amount of impurities added to an otherwise pure system keeps the
critical behaviour unchanged, affecting only the value of the
critical temperature. In contrast, higher concentration of
impurities are known to alter the nature of the transition
qualitatively. Noting that in our model free energy (\ref{freegen})
the effective critical temperature $\tilde T_c=T_c-2\lambda \phi^2 -
2\overline\lambda\phi$ is composition-dependent, it would be
interesting to explore possible connections between our results here
and the discussions in Ref.~\cite{chamati}.
 While constructing the form of $\mathcal F$
as given in (\ref{freegen}), we have coupled composition with the
mean curvature $H$, but have ignored the Gaussian curvature $G$. For
a pure fluid membrane with a fixed topology, $G$ does not play any
role in the statistical mechanics of the system, due to
the Gauss-Bonnet theorem~\cite{chaikin}. For a heterogeneous membrane, it is
possible to introduce terms coupling $\phi$ and $G$.
However, they do not contribute to the renormalisation
of $\kappa$ to the first order in $T/\kappa$, and hence, are neglected. We
have ignored the technical issue of choosing the correct measure in
$\mathcal Z$\cite{lub1}. These are not expected to affect our low
order perturbative results. Our model may be extended for
multicomponent (more than two components) heterogeneous membranes in a straight
forward way, by introducing additional composition variables and
coupling all of them to the mean curvature in ways similar to
(\ref{freegen1}). We expect that the general features of our results
should hold there. Lastly, dynamical behaviour of  a heterogeneous
membrane modelled by (\ref{freegen}) near $T_c$ would be interesting
to study theoretically (see, e.g., Ref.\cite{sunil} for a  study of
the dynamics of a heterogeneous membrane with a
curvature-composition interaction from (\ref{freegen})).

\section{Acknowledgement}
AB wishes to thank the Max-Planck-Gesellschaft (Germany) and
Department of Science and Technology/Indo-German Science and
Technology Centre (India) for partial financial support through the
Partner Group programme (2009).


\begin{thebibliography}{99}
\bibitem{renlam} There are evidently no corrections to $\lambda$ and $\overline\lambda$
at the lowest order in $O(T/\kappa)$.
\bibitem{phiren} We have ignored renormalisation of $\phi$ due to the
geometric nonlinearities, which would produce corrections $O(T/\kappa)$. Thus
to $O(T/\kappa)^0$ this may be neglected. See also Sec.~\ref{conclu} for
additional discussions.
\bibitem{1storder} In principle the generalised free energy functional (\ref{freegen1} may admit first order transitions in addition to the second order transition at $T=T_c$. We ignore these possibilities here.
\end{thebibliography}
\end{document}